\documentclass[aps,prd,groupaddress,tighten,nofootinbib,preprint,superscriptaddress]{revtex4}
\usepackage{amssymb}
\usepackage{amsmath,color}
\usepackage{epsfig}
\usepackage{graphicx,epsf,epsfig}
\usepackage{bbm}
\def\slc#1{\setbox0=\hbox{$#1$}           
    \dimen0=\wd0                                 
    \setbox1=\hbox{/} \dimen1=\wd1               
    \ifdim\dimen0>\dimen1                        
       \rlap{\hbox to \dimen0{\hfil/\hfil}}      
       #1                                        
    \else                                        
       \rlap{\hbox to \dimen1{\hfil$#1$\hfil}}   
       /                                         
    \fi}
\begin{document}
\preprint{NORDITA-2009-31}
\title{Non-unitary neutrino mixing and CP violation in the minimal inverse seesaw model}
\date{\today}
\author{Michal Malinsk\'y}
\email{malinsky@kth.se}

\author{Tommy Ohlsson}
\email{tommy@theophys.kth.se}

\affiliation{Department of Theoretical Physics, School of
Engineering Sciences, Royal Institute of Technology (KTH) --
AlbaNova University Center, Roslagstullsbacken 21, 106 91 Stockholm,
Sweden}

\author{Zhi-zhong Xing}
\email{xingzz@ihep.ac.cn}

\affiliation{Institute of High Energy Physics and Theoretical
Physics Center for Science Facilities, Chinese Academy of Sciences,
P.O. Box 918, Beijing 100049, China}

\author{He Zhang}
\email{zhanghe@kth.se}

\affiliation{Department of Theoretical Physics, School of
Engineering Sciences, Royal Institute of Technology (KTH) --
AlbaNova University Center, Roslagstullsbacken 21, 106 91 Stockholm,
Sweden}
\begin{abstract}
We propose a simplified version of the inverse seesaw model, in
which only two pairs of the gauge-singlet neutrinos are introduced,
to interpret the observed neutrino mass hierarchy and lepton flavor
mixing at or below the TeV scale. This ``minimal" inverse seesaw
scenario (MISS) is technically natural and experimentally testable.
In particular, we show that the effective parameters describing the
non-unitary neutrino mixing matrix are strongly correlated in the
MISS, and thus, their upper bounds can be constrained by current
experimental data in a more restrictive way. The Jarlskog invariants
of non-unitary CP violation are calculated, and the discovery
potential of such new CP-violating effects in the near detector of a
neutrino factory is discussed.
\end{abstract}
\maketitle

\section{Introduction}
\label{sec:intro}

Among various theoretical attempts, the famous seesaw ideas provide
us with a very natural way to understand why the masses of three
known neutrinos are so tiny compared to the masses of other Standard
Model (SM) fermions
\cite{Minkowski:1977sc,Yanagida:1979as,Mohapatra:1979ia,Schechter:1980gr,Lazarides:1980nt,Mohapatra:1980yp,Foot:1988aq}.
In the canonical type-I seesaw mechanism, in which three
right-handed (RH) neutrinos are introduced and lepton number
violation is allowed, the effective mass matrix of three light
Majorana neutrinos $m_\nu$ is dramatically suppressed with respect
to the electroweak scale if the RH neutrino mass matrix $M_{\rm R}$
is located not far away from the typical scale of grand unified
theories. As a rough estimate, if light neutrino masses are
stabilized around the sub-eV scale and the Dirac mass matrix $M_{\rm
D}$ between left- and right-handed neutrinos is comparable with the
mass of the top quark, then $M_{\rm R} \sim 10^{14} ~{\rm GeV}$ is
naturally expected. The testability of such conventional seesaw
models is therefore questionable. On the other hand, it is difficult
to embed RH neutrinos into a theoretical framework at low-energy
scales [e.g., the TeV scale to be explored by the Large Hadron
Collider (LHC)] in a technically natural way while keeping the
left-handed neutrinos to be light enough. For those realistically
viable type-I or type-III seesaw models with the TeV-scale RH
neutrinos, which could be experimentally accessible at the LHC,
fine-tunings of cancellations among the contributions to $m_\nu$
from different heavy neutrinos have to be employed. These kinds of
structural cancellations are usually attributed to some underlying
flavor symmetries
\cite{Buchmuller:1991tu,Pilaftsis:1991ug,Ingelman:1993ve,Heusch:1993qu,Kersten:2007vk}.
In the type-(I+II) seesaw model, one may also assume the mass term
coming from RH neutrinos to be comparable with the one coming from
the triplet Higgs, and thus, the light neutrino mass scale is
brought down through a significant cancellation between these two
mass terms \cite{Chao:2007mz}. However, to generate appreciable
collider signatures of the heavy seesaw particles at the LHC, such a
scheme potentially suffers from dangerous radiative corrections and
requires unnatural fine-tuning even at loop level
\cite{Chao:2008mq}.

In our recent work \cite{Malinsky:2009gw}, we have pointed out that
the above-mentioned drawbacks of most TeV-scale type-I, type-III, and
type-(I+II) seesaw models can be circumvented by considering the
inverse seesaw model \cite{Mohapatra:1986bd}. In the latter
framework, additional SM gauge singlets are adhibited together with
a small Majorana mass insertion which explicitly breaks the lepton
number. The phenomenology of this inverse seesaw mechanism is very
rich: on the one hand, non-unitary neutrino mixing and CP violation
can naturally show up and are possible to be tested at the future
long-baseline neutrino oscillation experiments; on the other hand,
the heavy seesaw particles may result in very attractive signatures
of lepton-flavor-violating (LFV) processes at the LHC. Since the
heavy singlets possess opposite CP signs and compose the
pseudo-Dirac particles in pair, the lepton-number-violating (LNV)
processes (such as the like-sign di-lepton events at the LHC) are
significantly suppressed and practically invisible.

However, the inverse seesaw scenario with three heavy singlet pairs
contains too many free parameters and is not very predictive. To
partly avoid this drawback, one may consider a simplified version of
the generic inverse seesaw scenario by reducing its number of
degrees of freedom. This sound motivation leads us to the minimal
inverse seesaw scenario (MISS) at the TeV scale, in which only two
pairs of the SM gauge-singlet neutrinos are introduced but the
observed neutrino mass hierarchy and lepton flavor mixing can well
be interpreted.\footnote{See also the minimal type-I seesaw
\cite{Frampton:2002qc,Guo:2006qa} and the minimal type-(I+II) seesaw
\cite{Gu:2006wj}.}

The purpose of this work is to describe the MISS, which contains
only two RH neutrinos ($\nu_{{\rm R}1}$, $\nu_{{\rm R}2}$) and two
SM gauge singlets ($S_1$, $S_2$), and to explore some of its
low-energy consequences on neutrino mixing and CP violation. The
remainder of our work is organized as follows. In
Sec.~\ref{sec:MISS}, we will introduce the MISS and present some
general formulas associated with the non-unitarity of the light
neutrino mixing matrix. In Sec.~\ref{sec:phenomenon}, we will
constrain the parameter space of non-unitarity effects in the MISS
by using current experimental data, calculate the Jarlskog
invariants of leptonic CP violation, and discuss the discovery
potential of such new CP-violating effects in the near detector of a
neutrino factory. Finally, a brief summary will be given in
Sec.~\ref{sec:summary}.

\section{The minimal inverse seesaw scenario}
\label{sec:MISS}

The MISS is constructed by extending the SM particle content with
two RH neutrinos $\nu_{\rm R} = (\nu_{\rm R1},\nu_{\rm R2})$ and two
left-handed (LH) SM gauge singlets $S=(S_1,S_2)$. The mass part of
the neutrino sector Lagrangian is then arranged so that it reads in
the flavor basis
\begin{eqnarray}\label{eq:Lmass}
-{\cal L}_{\rm m} = \overline{\nu_{\rm L}} M_{\rm D} \nu_{\rm R} +
\overline{S} M_{\rm R} \nu_{\rm R} + \frac{1}{2} \overline{S} \mu
S^c + {\rm H.c.} \ ,
\end{eqnarray}
where $\mu$ is a complex symmetric $2 \times 2$ matrix and $M_{\rm
D}$ and $M_{\rm R}$ are arbitrary $3\times2$ and $2\times2$
matrices, respectively. Without loss of generality, one can always
redefine the extra singlet fields and work in a basis where $\mu$ is
real and diagonal, namely, $\mu={\rm diag}(\mu_1,\mu_2)$ with
$\mu_1<\mu_2$. In addition to that, $M_{\rm R}$ can be made
Hermitian by a further unitary transformation in the RH neutrino
sector. The $7\times7$ neutrino mass matrix in the basis $(\nu_{\rm
L},\nu^c_{\rm R},S)$ is then rewritten as
\begin{eqnarray}\label{eq:Mnu}
M_{\nu}=\left(\begin{array}{ccc}
0 & M_{\rm D} & 0 \\
M^T_{\rm D} & 0 & M^T_{\rm R} \\
0 & M_{\rm R} & \mu
\end{array}\right) \ ,
\end{eqnarray}
which is clearly a symmetric matrix with rank at most 6 \cite{Xing:2007uq}.

Note that $M_{\rm R}$ is a SM singlet mass term, and hence, it is not
governed by the scale of the $SU(2)_{\rm L}$ symmetry breaking. In what
follows, we will consider a particularly attractive case $M_{\rm R}
> M_{\rm D} \gg \mu$, with $M_{\rm R}$ not far above
the electroweak scale. In the limit $\mu \to 0$, the rank of
$M_{\nu}$ reduces from 6 to 4, which leaves three light neutrinos
massless. In reality, a tiny but non-vanishing $\mu$ can be viewed
as a slight breaking of a global $U(1)$ symmetry, and thus, it
respects the naturalness criterion \cite{tHooft:1979bh}. In contrast
to the original inverse seesaw scenario, the MISS predicts one light
neutrino to be exactly massless and brings in several interesting
phenomena in neutrino oscillations and LFV processes.

At the leading order in $M_{\rm
D}M^{-1}_{\rm R}$, the light neutrino mass matrix in the MISS is
given by
\begin{eqnarray}\label{eq:mnu}
m_{\nu}\simeq M_{\rm D} M^{-1}_{\rm R} \mu (M^T_{\rm R})^{-1} M_{\rm
D}^{T}  \equiv F \mu F^T\ ,
\end{eqnarray}
where $F=M_{\rm D} M^{-1}_{\rm R}$ is a $3\times 2$ matrix. For
$\mu$ around the keV scale, $M_{\rm D}M^{-1}_{\rm R} \sim 10^{-2}$
gives rise to the desired sub-eV light neutrino masses. In the limit
$\mu \rightarrow 0$, we have $m_\nu = 0$, which corresponds to the
lepton number symmetry restoration. The heavy sector consists of a
pair of pseudo-Dirac neutrinos $P_j=(S_j,\nu_{{\rm R}j})$
\cite{Bilenky:1987ty} with a tiny mass splitting between the
relevant CP-conjugated Majorana components of the order of $\mu$.

One can diagonalize $m_{\nu}$ by means of a unitary transformation
\begin{eqnarray}\label{eq:mnuV}
U^\dagger m_{\nu} U^* = \bar{m}_\nu ={\rm diag} (m_1,m_2,m_3) \ ,
\end{eqnarray}
with $m_i$ (for $i=1,2,3$) denoting the light neutrino masses. We are left with either
$m_1=0$ (normal mass hierarchy $\Delta m^2_{31} >0$, NH) or $m_3=0$
(inverted mass hierarchy $\Delta m^2_{31} <0$, IH). Note that $U$
itself is not the matrix that governs neutrino
oscillations even if we choose a basis where the charged-lepton mass
matrix is diagonal.

The LH neutrinos entering the charged-current interactions of the SM
are superpositions of the seven mass eigenstates $(\nu_{m{\rm
L}},P_m)$ given at the leading order by
\begin{eqnarray}
\nu_{\rm L} \simeq N \nu_{m{\rm L}} + F U_{\rm R} P_{m\rm L}\ ,
\end{eqnarray}
where $U^\dagger_{\rm R} M_{\rm R} U_{\rm R} = {\rm diag}
(m^{}_{P_1}, m^{}_{P_2})$ and $N \simeq
\left(1-\frac{1}{2}FF^\dagger\right) U $ \cite{Schechter:1981cv}.
Hence, one can write
\begin{eqnarray}
{\cal L}_{\rm CC} = -\frac{g}{\sqrt{2}}  W^-_\mu \overline{\ell_{\rm
L}} \gamma^\mu \left( N \nu_{m \rm L} +F U_{\rm R} P_{m\rm L}\right)
+ {\rm H.c.}
\end{eqnarray}
The mixing between the doublet and singlet components in the charged
currents results in several interesting phenomenological
consequences:
\begin{itemize}
\item The flavor and mass
eigenstates of the left-handed neutrinos are connected by a
non-unitary flavor mixing matrix $N$ \cite{Antusch:2006vwa}. The
magnitudes of non-unitarity effects in different neutrino
oscillation channels are predominated by the mass ratios between
$M_{\rm D}$ and $M_{\rm R}$, and in principle, their underlying
correlations have to be taken into account in analyses of the future
experiments.

\item The heavy singlets entering the charged currents due to the
non-unitarity effects also enter the rare lepton decays, such as
$\tau \to \mu \gamma$ and $\mu \to e \gamma$. Hence, unlike in the
type-I seesaw model, their contributions to the LFV decays are not
suppressed by the light neutrino masses \cite{Deppisch:2004fa}, but
mainly constrained by the ratio $M_{\rm D}M^{-1}_{\rm R}$, which in
principle admits observing these events in the upcoming LHC
experiments. For example, the decays $\ell_\alpha \rightarrow
\ell_\beta \gamma$ are mediated by $P$'s and their branching ratios
are given by \cite{Ilakovac:1994kj}
\begin{eqnarray}\label{eq:B}
{\rm BR}\left({\ell_\alpha \rightarrow \ell_\beta \gamma}\right) =
\frac{\alpha^3_W s^2_W m^5_{\ell_\alpha}}{256 \pi^2 M^4_W
\Gamma_\alpha }  \left| \sum^2_{i=1} K_{\alpha i}K_{\beta i}^{*}
I\left(\frac{m^{}_{P_i}}{M^2_W}\right) \right|^2 \ ,
\end{eqnarray}
where $K=FU_{\rm R}$, $I\left(x\right)= -(2x^3+5x^2-x)/[4(1-x)^3] -
3x^3\ln x/[2(1-x)^4]$, and $\Gamma_\alpha$ is the total width of
$\ell_\alpha$. In the conventional type-I seesaw model (i.e.,
without unnatural cancellations), one has approximately
$KK^{\dagger}= {\cal O}(m_{\nu}M_{\rm R}^{-1})$, and therefore,
${\rm BR}\left({\ell_\alpha \rightarrow \ell_\beta \gamma}\right)
\propto {\cal O} (m_{\nu}^{2})$ indicates a strong suppression of
LFV decays. However, in the inverse seesaw model, one can have
sizable $K$ without any reference to the tiny
neutrino masses, since the two issues are essentially decoupled. Thus,
appreciable LFV rates could be obtained even for strictly massless
light neutrinos \cite{Bernabeu:1987gr}.

\item
If the masses of the heavy singlets $P_{m}$ do not fall far beyond
the electroweak scale, in the MISS, as in the type-I seesaw model,
one can expect an on-shell production of $P$ at the LHC via the
gauge boson exchange diagrams. The most distinctive signature would
be the observation of LFV processes involving three charged-leptons
in the final state \cite{delAguila:2008hw}.
\end{itemize}

In comparison to the generic inverse seesaw scenario, the MISS is
certainly more restrictive and predictive because of its much fewer
free parameters and underlying correlations among physical
observables, which could be well tested at a future neutrino
factory. In the remaining part of this work, we will discuss the
phenomenological consequences mentioned above in more detail and, in
particular, concentrate on the possible non-unitary CP-violating
effects in neutrino oscillations.

\section{Non-unitarity effects} \label{sec:phenomenon}

\subsection{Constraints on non-unitarity parameters}

As we have already mentioned, light neutrino mass eigenstates are
connected to their flavor eigenstates by a non-unitary mixing matrix
$N$. In terms of the parametrization advocated in
Ref.~\cite{Altarelli:2008yr}, the non-unitary leptonic mixing is
written as $N=(1-\eta)U$ where the relevant Hermitian matrix $\eta$
is given by $ \eta \simeq \frac{1}{2} FF^\dagger $ and the unitary
matrix $U$ can be parametrized in the standard form
\begin{eqnarray}\label{eq:para}
U & = &\left(
\begin{matrix}c_{12} c_{13} & s_{12} c_{13} & s_{13}
e^{-{\rm i}\delta} \cr -s_{12} c_{23}-c_{12} s_{23} s_{13}
 e^{{\rm i} \delta} & c_{12} c_{23}-s_{12} s_{23} s_{13}
 e^{{\rm i} \delta} & s_{23} c_{13} \cr
 s_{12} s_{23}-c_{12} c_{23} s_{13}
 e^{{\rm i} \delta} & -c_{12} s_{23}-s_{12} c_{23} s_{13}
 e^{{\rm i} \delta} & c_{23} c_{13}\end{matrix}
\right) \left(
\begin{matrix} 1 & & \cr  & e^{{\rm i}\rho} & \cr & & 1 \end{matrix}
\right) \ ,
\end{eqnarray}
provided $c_{ij} \equiv \cos \theta_{ij}$ and $s_{ij} \equiv \sin
\theta_{ij}$ (for $ij=12$, $13$, $23$). Since one light neutrino
mass is vanishing in the MISS, we have only one Majorana phase. The
current experimental bounds on $\eta$ are rather stringent, implying
that one can approximate $\theta_{ij}$'s by the values of the mixing
angles obtained from the neutrino oscillations. Present data on
atmospheric, solar, and reactor neutrinos yield two neutrino
mass-squared differences $\Delta m^2_{21} \simeq 7.65 \times
10^{-5}~{\rm eV}$ and $|\Delta m^2_{31} | \simeq 2.40 \times
10^{-3}~{\rm eV}$, together with three neutrino mixings
$\sin^2\theta_{12} \simeq 0.304$, $\sin^2\theta_{23} \simeq 0.50$,
and $\sin^2\theta_{13} \simeq 0.01$ \cite{Schwetz:2008er}. There is
no hint on the CP-violating phases, and thus, we take
$\theta_{13}<10^\circ$ and leave $\delta$ as a free parameter in the
following calculations. As for the non-unitarity parameters,
$|\eta_{\alpha\beta}|$ are constrained mainly from universality
tests of weak interactions, rare leptonic decays, invisible width of
the $Z$-boson and neutrino oscillation data. The present bounds on
$|\eta_{\alpha\beta}|$ (at the 90~\% C.L.) are
\cite{Antusch:2008tz}:
\begin{eqnarray}\label{eq:etaB}
|\eta| \equiv (|\eta_{\alpha\beta}|) < \left(\begin{matrix}
2.0\times10^{-3} & 6.0 \times 10^{-5} & 1.6\times10^{-3} \cr \sim &
8.0\times10^{-4} & 1.1 \times10^{-3} \cr \sim & \sim &
2.7\times10^{-3}
\end{matrix}\right)\ .
\end{eqnarray}

In order to study the connections among physical parameters in the
MISS, we adopt the parametrization of $F$ from the work in
Ref.~\cite{Ibarra:2003up}, namely
\begin{eqnarray}\label{eq:F}
F = U \sqrt{\bar{m}_\nu} R \sqrt{\mu^{-1}} \ ,
\end{eqnarray}
where
\begin{eqnarray}\label{eq:RNH}
R =  \left(\begin{matrix} 0 & 0 \cr \cos z & -\sin z \cr \sin z &
\cos z \end{matrix}\right) \ ,
\end{eqnarray}
for the NH case, and
\begin{eqnarray}\label{eq:RIH}
R =  \left(\begin{matrix}  \cos z & -\sin z \cr \sin z & \cos z \cr
0 & 0
\end{matrix}\right) \ ,
\end{eqnarray}
for the IH case. Here $z=\alpha + {\rm i}\beta$ is an arbitrary
complex number with both $\alpha$ and $\beta$ being real. For sake
of simplicity, in what follows we will use the parameters
$r\equiv\mu_1/\mu_2$ and $\epsilon \equiv \sqrt[4]{\Delta
m^2_{21}/|\Delta m^2_{31}|} \simeq 0.42$. In the NH case, we obtain
\begin{eqnarray}\label{eq:FNH}
F = \sqrt{\frac{m_3}{\mu_2}}\left(\begin{matrix} 0 & 0 \cr
\frac{1}{\sqrt{r}} s_{23} s_z & s_{23} c_z \cr \frac{1}{\sqrt{r}}
c_{23} s_z & c_{23} c_z
\end{matrix}\right) + {\cal O}(\epsilon) \ ,
\end{eqnarray}
which yields
\begin{eqnarray}\label{eq:etaNH}
\eta \simeq \frac{1}{2} \frac{m_3}{\mu_2}
\frac{r|c_z|^2+|s_z|^2}{2r} \left(\begin{matrix} 0 & 0 & 0 \cr 0 &
s^2_{23} & s_{23} c_{23} \cr 0 & s_{23} c_{23} & c^2_{23}
\end{matrix}\right) \ .
\end{eqnarray}
Thus, the only non-negligible off-diagonal entry is
$\eta_{\mu\tau}$, and we have the relation $|\eta_{\mu\mu}|\simeq|
\eta_{\mu\tau}|\simeq|\eta_{\tau\tau}|$ because, to a good approximation,
$\theta_{23} \simeq 45^\circ$.
Similarly, in the IH case, we
obtain
\begin{eqnarray}\label{eq:FIH}
F = \sqrt{\frac{m_1}{\mu_2}} \left(\begin{matrix} \frac{1}{\sqrt{r}}
A & B \cr - \frac{1}{\sqrt{r}} c_{23} X & c_{23} Y \cr
\frac{1}{\sqrt{r}} s_{23} X & -s_{23} Y \end{matrix}\right) + {\cal
O}(\epsilon^2) \ ,
\end{eqnarray}
which translates into
\begin{eqnarray}\label{eq:etaIH}
\eta \simeq \frac{1}{2} \frac{m_1}{\mu_2} \left(\begin{matrix}
\frac{1}{r}A^2+B^2 & -c_{23} \left( \frac{1}{r}AX^* - BY^* \right) &
s_{23} \left( \frac{1}{r}AX^* - BY^* \right) \cr \sim & c^2_{23}
\left(\frac{1}{r} |X|^2 +|Y|^2 \right)  & -s_{23} c_{23}
\left(\frac{1}{r} |X|^2 +|Y|^2 \right)  \cr \sim &\sim & s^2_{23}
\left(\frac{1}{r} |X|^2 +|Y|^2 \right)
\end{matrix}\right) \ ,
\end{eqnarray}
where $A=c_{12}c_z+e^{{\rm i}\rho} s_{12}s_z$, $B=-c_{12}s_z+e^{{\rm
i}\rho} s_{12}c_z$, $X=s_{12}c_z - e^{{\rm i}\rho} c_{12}s_z$, and
$Y=s_{12}s_z+e^{{\rm i}\rho} c_{12}c_z$. Hence, in both cases, the
approximately maximal leptonic 23 mixing implies
$|\eta_{\mu\mu}|\simeq |\eta_{\mu\tau}|\simeq|\eta_{\tau\tau}|$ and
$|\eta_{e\mu}|\simeq |\eta_{e\tau}|$.
\begin{figure}[th]
\begin{center}\vspace{-1.cm}
\includegraphics[width=16cm]{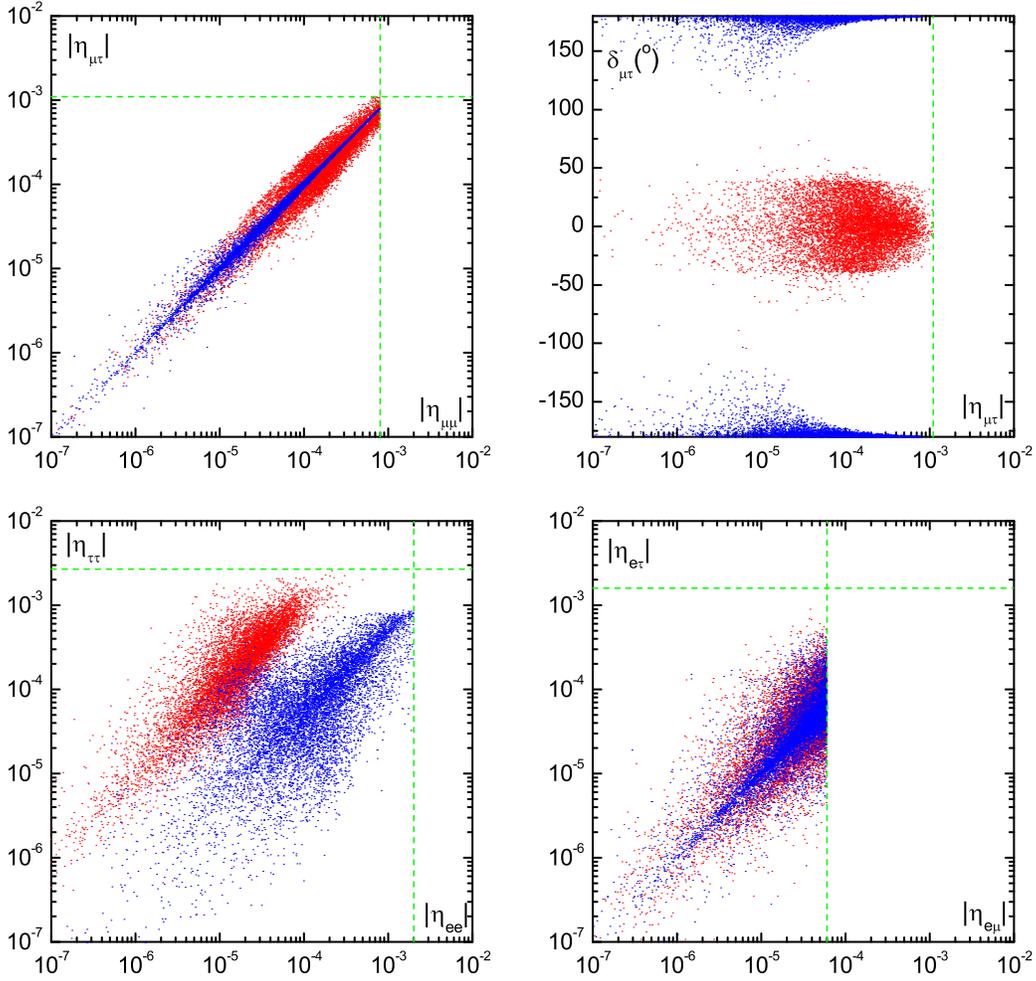}\vspace{-1.cm}
\caption{\label{fig:eta} Correlations among various parameters
governing the non-unitarity effects in the MISS. Red points denote
the normal neutrino mass hierarchy, while blue points correspond to
the case of the inverted mass hierarchy. Generic experimental
constraints are indicated by the green dashed lines (and, for
simplicity, any would-be correlations in their determination have
been neglected).} \vspace{-0.cm}
\end{center}
\end{figure}

\begin{figure}[h]
\begin{center}\vspace{0.cm}
\includegraphics[width=7.5cm]{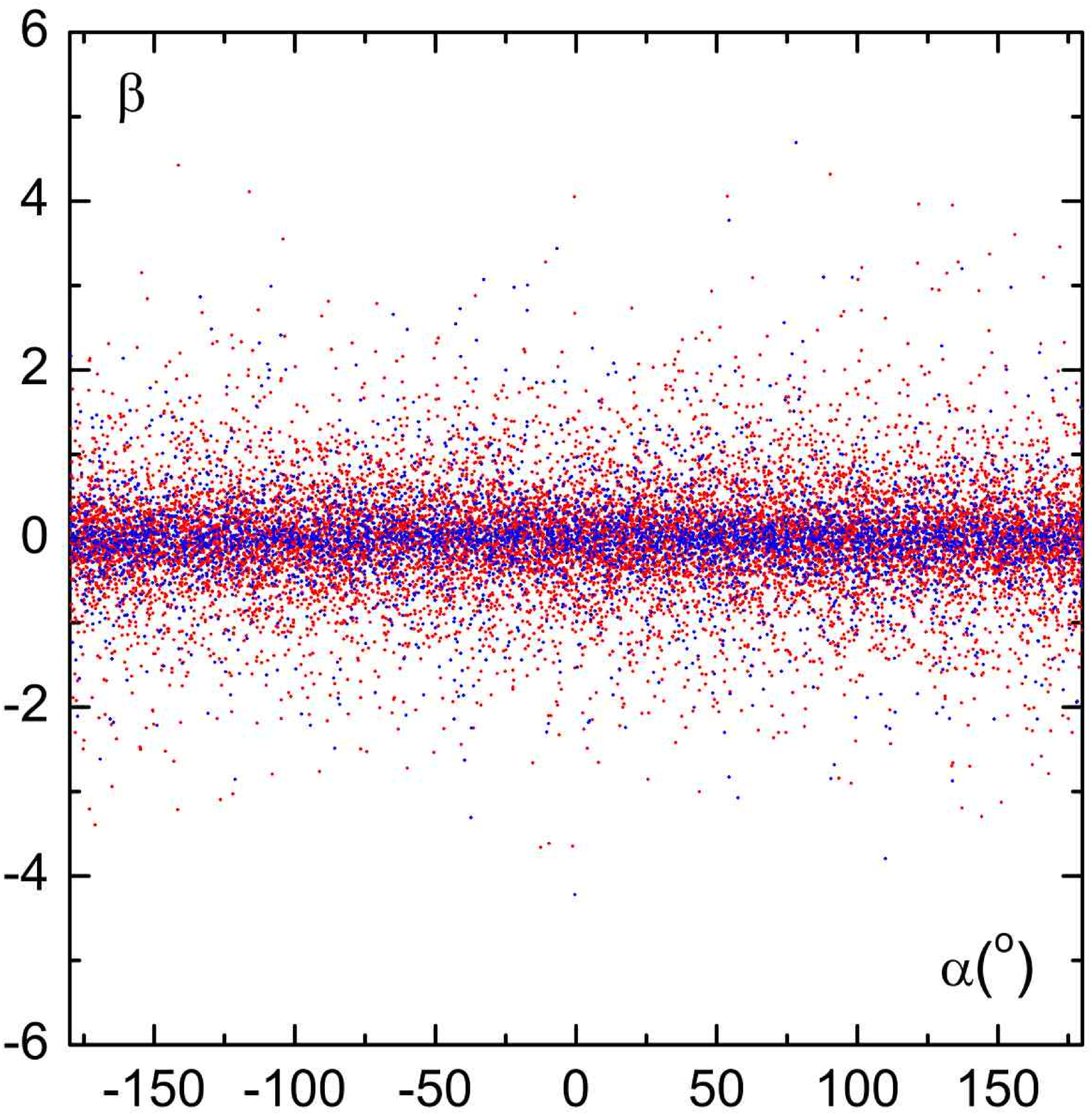}
\includegraphics[width=7.5cm]{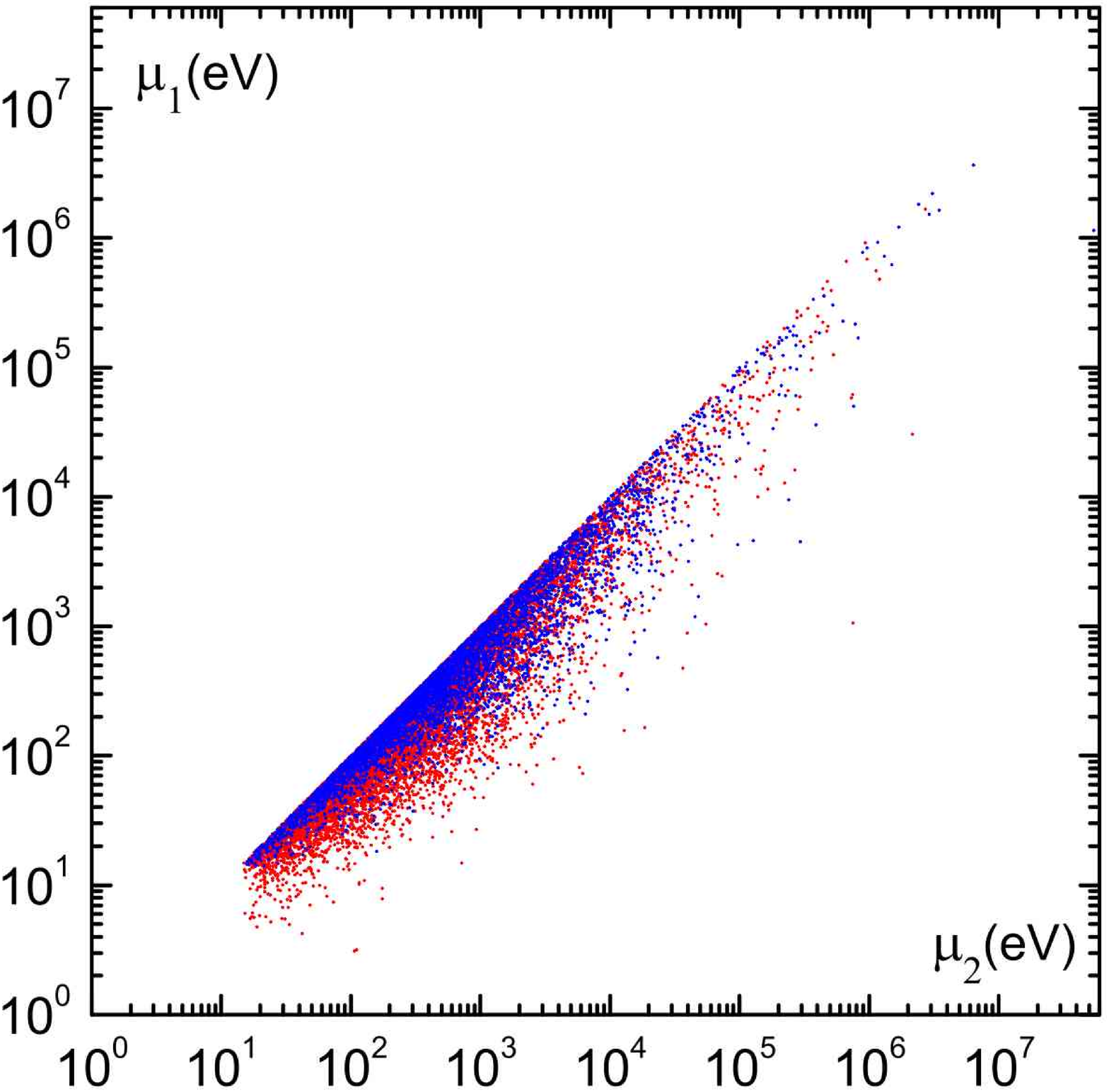}
\caption{\label{fig:mu} Constraints on the underlying model
parameters. As before, red and blue points denote the normal and
inverted neutrino mass hierarchies, respectively. As described in
the text, the $|\beta|\gg 1$ and $\mu_{2}\to 0$ regions correspond
to a fine-tuning, which we wish to avoid, and thus, the scanning
granularity is large. Moreover, $\mu_{1}\to 0$ leads to a further
reduction of the rank of $m_{\nu}$, and hence, it is also
disfavored.} \vspace{-0.cm}
\end{center}
\end{figure}

The allowed regions of non-unitarity parameters are illustrated in
Figs.~\ref{fig:eta} and \ref{fig:mu}. In total, we have eleven
parameters out of which seven
($\mu_1,\mu_2,\alpha,\beta,\rho,\delta,\theta_{13}$) are essentially
free while the remaining four ($\theta_{12},\theta_{23},\Delta
m_{21}^2,\Delta m_{31}^2$) are quantities fixed by their best-fit
values. In our numerical analysis, we randomly generate points in
the seven-dimensional parameter space. The points that are within
the 90~\% C.L.~upper bounds on the non-unitarity parameters
$\eta_{\alpha\beta}$ are plotted in the figures, where we do not
resort to the approximate Eqs.~(\ref{eq:FNH})-(\ref{eq:etaIH}) but
rather utilize the precise Eq.~(\ref{eq:F}). As a result, $10^4$
points build up each plot. Using Eq.~\eqref{eq:mnu}, one can easily
estimate that $\mu_2 \gg m_i$ should be ensured if there is no
strong structural cancellations in $F$. Hence, in our numerical
calculations, we set $|\Delta m^2_{31}|/\mu_2 < 0.01$ as a prior in
order to avoid unnatural fine-tuning among model parameters.

It is important to make clear that the decrease of the point-density
towards smaller values of $|\eta_{\alpha\beta}|$ is a mere numerical
artifact, since these regions correspond to a physically not very
interesting situation, and hence, it does not make sense to scan
over such areas thoroughly. On the other hand, the density reduction
observed in the opposite (i.e., growing $|\eta_{\alpha\beta}|$)
directions, c.f., the two upper plots in Fig.~\ref{fig:eta}, provides
a true physical information and all the following statements are
based on such kind of physically relevant features.

In Fig.~\ref{fig:eta}, we present the allowed regions for the
absolute values of the relevant non-unitarity parameters. The
upper-left plot shows the correlation between $|\eta_{\mu\mu}|$ and
$|\eta_{\mu\tau}|$, which is stronger in the IH case than in the NH
case. This is expected, since the analytical approximation in the IH
case is better than that in the NH case. The upper-right plot shows
the CP-violating phase $\delta_{\mu\tau}$ of the parameter
$\eta_{\mu\tau}$, which is centered around 0 in the NH case and
around $\pm 180^\circ$ in the IH case, in agreement with
Eqs.~\eqref{eq:etaNH} and \eqref{eq:etaIH}. Thus, in both cases, it
is hard to achieve a sizable $|\eta_{\mu\tau}|$ and maximal
CP-violating effects simultaneously. The bounds on the other $\eta$
parameters are shown in the plots in the second row of
Fig.~\ref{fig:eta}. One can observe that $|\eta_{ee}|$ is mainly
constrained in the NH case, while $|\eta_{\tau\tau}|$ is restricted
in the IH case. The allowed region for $|\eta_{e\tau}|$ is limited
in both NH and IH cases, and there is no upgraded bound on
$|\eta_{e\mu}|$. Note that, in the NH case, the leading order
formulae \eqref{eq:FNH} and \eqref{eq:etaNH} do not provide as good
approximation of the precise results as do the Eqs.~\eqref{eq:FIH}
and \eqref{eq:etaIH} in the IH case, because the relevant effective
expansion parameters (i.e., $\varepsilon^{2}$ in the latter whilst
only $\varepsilon$ in the former case) are not the same. This also
imprints into the different widths of the red and blue bands in
Fig.~\ref{fig:eta}.

The allowed regions for $\mu_i$ and $z$ are shown in
Fig.~\ref{fig:mu}. There is no strong constraint on $\alpha$.
However, $\beta$ is bounded, since a larger $\beta$ corresponds to a
more sever fine-tuning of the model parameters. As for the $\mu_i$
parameters, there are no generic upper bounds one should impose;
however, smaller values of $\mu_i$ correspond to a stronger
fine-tuning entangled in $F$.

Finally, let us summarize the upgraded bounds on the non-unitarity
parameters we have obtained in the MISS under consideration:
$|\eta_{ee}| \lesssim 5.0 \times 10^{-4}$, $|\eta_{e\tau}| \lesssim
8.9\times 10^{-4}$ in the NH case, and $|\eta_{e\tau}| \lesssim
4.6\times 10^{-4}$, $|\eta_{\mu\tau}| \lesssim 7.9\times 10^{-4}$,
$|\eta_{\tau\tau}| \lesssim 8.8\times 10^{-4}$ in the IH case. They
could be helpful when building a specific and realistic model based
on the MISS.

\subsection{Jarlskog invariants \label{sec:jarlskoginas}}

In general, there are nine independent rephasing-invariant
quantities that one can build at the quartic level out of the
entries of a generic $3\times 3$ lepton mixing matrix $V$
\cite{Jarlskog:1985ht},
\begin{eqnarray}
J_{\alpha\beta}^{ij}={\rm Im}(V_{\alpha i}V_{\beta j}V^{*}_{\alpha
j}V^{*}_{\beta i})\ ,
\end{eqnarray}
where the indices $\alpha\neq \beta$ run over $e\mu$, $\mu\tau$ and
$\tau e$, while $i\neq j$ can be $12$, $23$ and $31$. Note that all
nine $J_{ab}^{ij}$'s coincide if $V$ is a unitary matrix, since all
six unitarity triangles, despite their different shapes, span the
same area \cite{Fritzsch:1999ee,Zhang:2004hf}. However, if
non-unitarity effects (like those studied in this work, i.e., $V =
N$) are present, this is no longer the case and one can expect
deviations from such a simple picture driven by the relevant
non-unitarity parameters (denoted by $\eta_{\alpha\beta}$ in the
current study). In such a case, it is instructive to know which
configuration of $\alpha\neq \beta$ and $i\neq j$ is  most affected
for a specific non-unitarity pattern. In particular, to leading
order in $\eta_{\alpha\beta}$, one can write
\begin{eqnarray}\label{eq:Jijab}
J_{\alpha\beta}^{ij} \simeq J + \Delta J_{\alpha\beta}^{ij} \ ,
\end{eqnarray}
where $J=c_{12} c_{13}^2 c_{23} s_{12} s_{13} s_{23} \sin\delta$
governs the CP-violating effects in the unitary limit. The second
term in Eq.~\eqref{eq:Jijab} depends on the off-diagonal (generally
complex) $\eta$'s, and hence, it does not necessarily vanish in the
limit $\theta_{13} \to 0$ and might even dominate the CP-violating
effects. The complete expressions for $\Delta J_{\alpha\beta}^{ij} $
are listed in Appendix \ref{app}. Focusing on the dominant
off-diagonal entry $|\eta_{\mu\tau}|$, c.f. Fig.~\ref{fig:eta}, the
following two contributions survive for $\eta_{e\mu}\to 0$ and
$\theta_{13}\to 0$:
\begin{eqnarray}
\Delta J_{\mu\tau}^{23} & = &
-|\eta_{\mu\tau}|\sin\delta_{\mu\tau}\sin2\theta_{23}\cos^{2}\theta_{12}
- |\eta_{e\tau}| s_{12} c_{12} s_{23}c^2_{23} \sin\delta_{e\tau} \ ,
\\
\Delta J_{\mu\tau}^{31} & = &
|\eta_{\mu\tau}|\sin\delta_{\mu\tau}\sin2\theta_{23}\sin^{2}\theta_{12}
- |\eta_{e\tau}| s_{12} c_{12} s_{23}c^2_{23}  \sin\delta_{e\tau}  \
,
\end{eqnarray}
provided $\eta_{\alpha\beta}=|\eta_{\alpha\beta}|e^{{\rm
i}\delta_{\alpha\beta}}$. Even beyond the simple limit above, one
can observe another interesting feature:
\begin{eqnarray}
J_{e\mu}^{23}= J_{e\mu}^{31} = J_{\tau e}^{23}=J_{\tau e}^{31}\ ,
\end{eqnarray}
where all the small parameters ($\eta_{\alpha\beta}$ and
$s_{13}$) have been kept at linear order. However, these
relations are mere reflections of the smallness of the mixing angle
$\theta_{13}$.

\begin{figure}[t]
\begin{center}\vspace{-0.5cm}
\includegraphics[width=7.5cm]{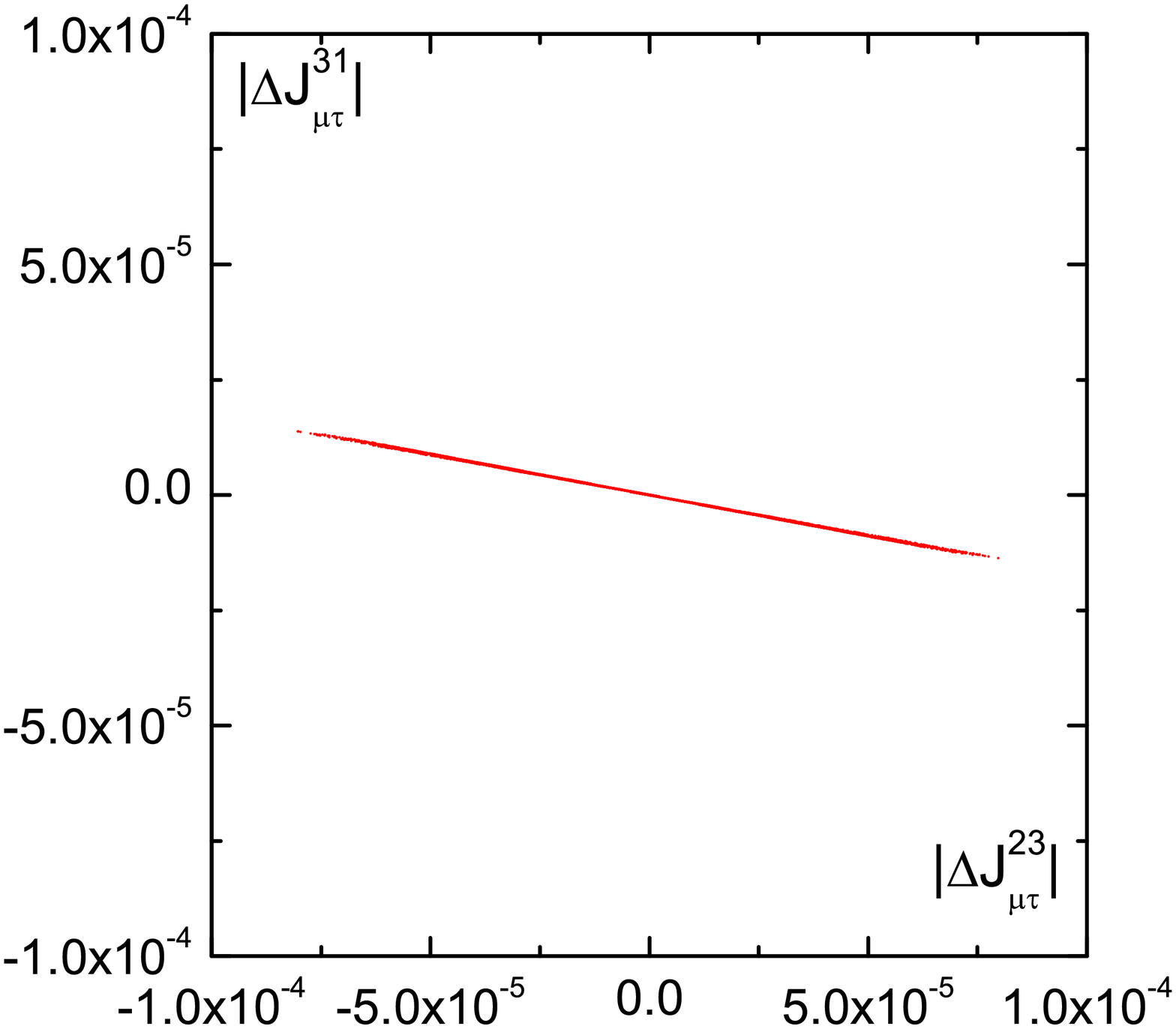}
\includegraphics[width=7.5cm]{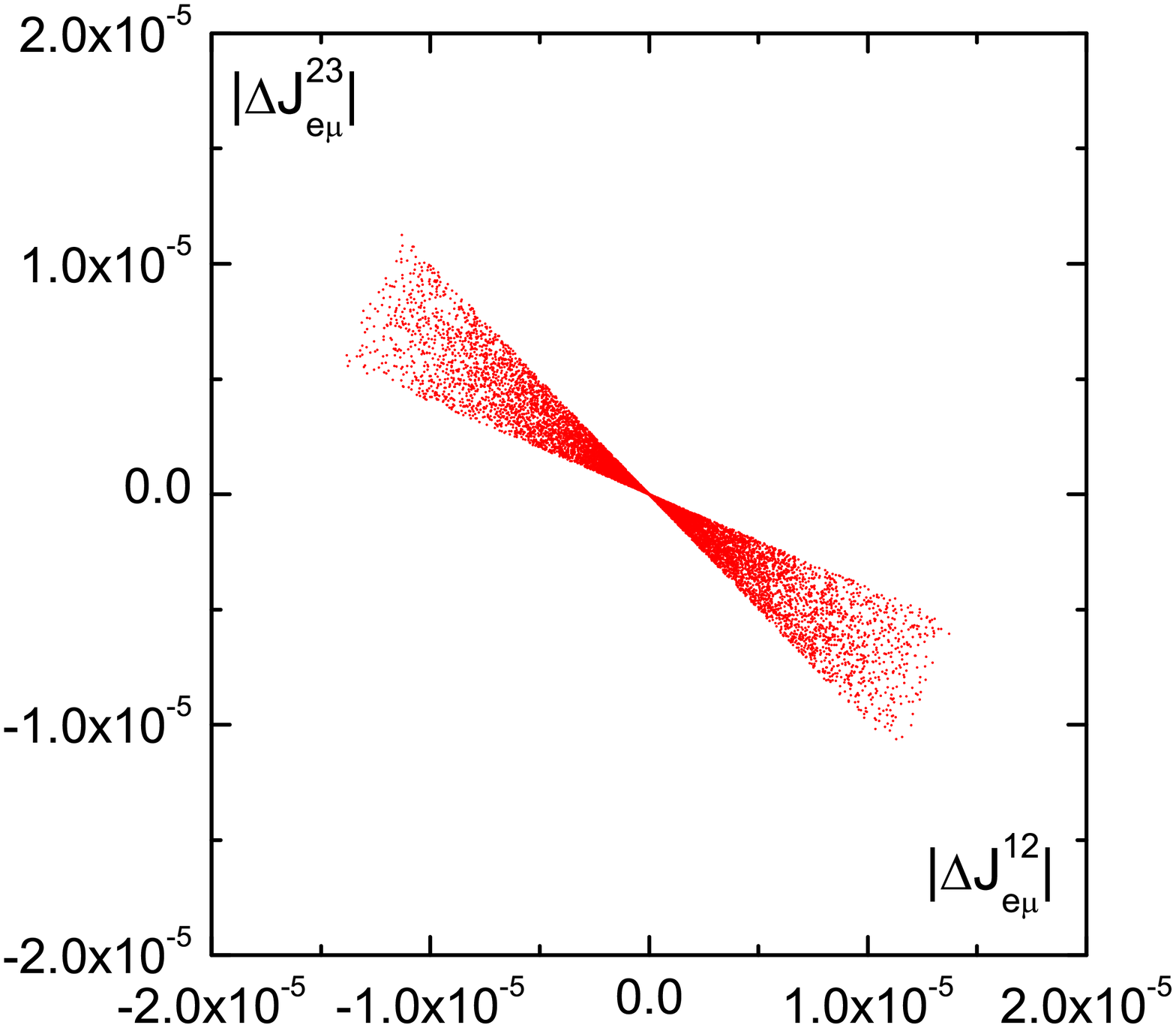}
\caption{\label{fig:J} Constraints on the Jarlskog invariants. Here,
only the NH case is considered, since the CP-violating effects due
to the non-unitarity parameters are small in the IH case. It is so
partly due to the smallness of $\eta_{e\tau}$ and also because of
the preferred value of the $\delta_{\mu\tau}$ phase, c.f.
Fig.~\ref{fig:eta}.} \vspace{-0.cm}
\end{center}
\end{figure}

In Fig.~\ref{fig:J}, we illustrate the correlations between Jarlskog
invariants in the NH case. As we expect, they are linearly
dependent, and the spread (i.e., the finite width of the allowed
strips) is due to higher-order corrections. Although the magnitude
of $\Delta J_{\mu\tau}^{31}$ does not seem to be comparable to that
of $J$, we will show later that the CP-violating effects induced by
the phases of non-unitarity parameters can be quite significant,
since they are not suppressed by the small neutrino mass-squared
difference $\Delta m^2_{21}$. For the IH case, according to
Eq.~\eqref{eq:etaIH}, $\eta_{e\mu} \sim \eta_{e\tau}$ are small
quantities, and the phase of $\eta_{\mu\tau}$ is close to $\pi$,
c.f. Fig.~\ref{fig:eta}. Hence, there is no observable CP-violating
effect coming from the non-unitarity parameters in the IH case.

\subsection{Sensitivity search at a neutrino factory}

For a non-unitary lepton flavor mixing matrix $N$, the vacuum
neutrino oscillation transition probability $P_{\alpha\beta}$ can be
written as \cite{Ohlsson:2008gx}
\begin{eqnarray}\label{eq:P}
P_{\alpha\beta} &=&  \sum_{i,j} {\cal F}^i_{\alpha\beta} {\cal
F}^{j*}_{\alpha\beta} - 4 \sum_{i>j} {\rm Re} ({\cal
F}^i_{\alpha\beta} {\cal F}^{j*}_{\alpha\beta} )\sin^2\!
\frac{\Delta m^{2}_{ij}L}{4E} +  2 \sum_{i>j}{\rm Im} ( {\cal
F}^i_{\alpha\beta} {\cal F}^{j*}_{\alpha\beta} ) \sin\frac{ \Delta
m^{2}_{ij} L}{2 E} \ , ~~~~~~
\end{eqnarray}
where $\Delta m^{2}_{ij} \equiv m^2_i - m^2_j$ are the neutrino
mass-squared differences and ${\cal F}^i$ are defined by
\begin{eqnarray}\label{eq:Fab}
{\cal F}^i_{\alpha\beta} \equiv \sum_{\gamma ,\rho} ( R^*)_{\alpha
\gamma } ( R^*)^{-1}_{\rho \beta } U^*_{\gamma i} U_{\rho i} \
\end{eqnarray}
with the normalized non-unitary factor
\begin{eqnarray}\label{eq:R}
R_{\alpha\beta} \equiv \frac{(1-\eta)_{\alpha\beta}}
{\left[(1-\eta)(1-\eta^\dagger)\right]_{\alpha\alpha}} \ .
\end{eqnarray}
If Earth matter effects are taken into account, then one can replace
the vacuum quantities $U$ and $m_i$ by their effective matter
counterparts, see e.g.~Ref.~\cite{Meloni:2009ia}.

As mentioned in the literature
\cite{FernandezMartinez:2007ms,Xing:2007zj,Goswami:2008mi,Donini:2008wz,Malinsky:2008qn,Tang:2009na},
the $\nu_\mu \rightarrow \nu_\tau$ channel together with a near
detector located at a short distance provides the most favorable
setup to constrain the non-unitarity effects.\footnote{An
alternative study for the disappearance channel $\nu_\mu \to
\nu_\mu$ together with a far detector located at $7500~{\rm km}$ has
been performed in Refs.~\cite{Blennow:2009pk,Antusch:2009pm}. In
particular, a sensitivity of ${\cal O}(10^{-4})$ could be achieved
due to matter effects, for which the neutrino oscillation
probability is only linearly suppressed in $\eta_{\mu\tau}$.}
In this respect, we consider the transition probability
$P_{\mu\tau}$ for a neutrino factory with a sufficiently short
baseline length $L$. We neglect the tiny matter effects and small
contributions of $\theta_{13}$ and $\Delta m^{2}_{21}$. Then,
$P_{\mu\tau}$ reads \cite{FernandezMartinez:2007ms}
\begin{eqnarray}\label{eq:Pmutau}
P_{\mu\tau} &\simeq & 4s^2_{23}c^2_{23} \sin^2\left(\frac{\Delta
m^{2}_{31} L}{4E}\right) - 4|\eta_{\mu\tau}|\sin\delta_{\mu\tau}
s_{23}c_{23} \sin\left(\frac{\Delta m^{2}_{31} L}{2E}\right) +
4|\eta_{\mu\tau}|^2 \ ,
\end{eqnarray}
where $E$ is the neutrino beam energy and the second term is CP-odd
due to the phase $\delta_{\mu\tau}$, and hence, distinctive
CP-violating effects can appear in neutrino oscillations
\cite{Altarelli:2008yr,Rodejohann:2009cq}. The last term in
Eq.~(\ref{eq:Pmutau}) plays the dominant role at `zero' distance, as
it does not depend on $L$.

Note that the shape of the CP-odd term is justified by the structure
of the relevant Jarlskog invariants derived in
Sec.~\ref{sec:jarlskoginas}. Indeed, the ${\rm Im} ( {\cal
F}^i_{\alpha\beta} {\cal F}^{j*}_{\alpha\beta} )$ factors in
Eq.~\eqref{eq:P} correspond to $-J_{\alpha\beta}^{ij}$ [apart from
the irrelevant real rescaling of $J$ due to the denominator in
Eq.~\eqref{eq:R}]. This means that the sum over $i>j$ in the last
term of Eq.~(\ref{eq:P}) is proportional to
$J_{\mu\tau}^{31}-J_{\mu\tau}^{23}=|\eta_{\mu\tau}|\sin\delta_{\mu\tau}
\sin 2\theta_{23}$ provided $\Delta m^{2}_{31} \simeq -\Delta
m^{2}_{23}$.

In order to show the feasibility of observing such a signal in the
future long-baseline neutrino oscillation experiments, we consider a
typical neutrino factory setting with an OPERA-like near detector with
fiducial mass of 5 kt. We assume a setup with approximately $
10^{21}$ useful muon decays and five years of neutrino running and
another five years of anti-neutrino running. We make use of the
GLoBES package \cite{Huber:2004ka,Huber:2007ji} with a slight
modification of the template Abstract Experiment Definition Language
(AEDL) file for the neutrino factory experiments
\cite{Autiero:2003fu,Huber:2006wb}.
\begin{figure}[th]
\begin{center}
\includegraphics[width=10cm]{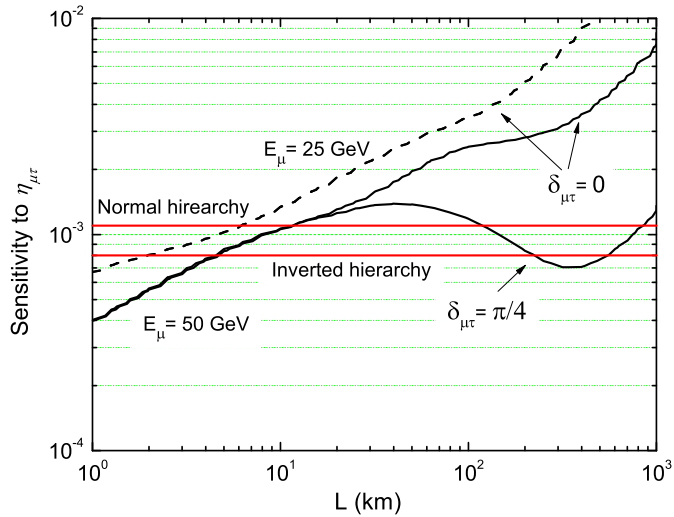}
\caption{\label{fig:NF} Sensitivity limits at $90~\%$ C.L.~on the
non-unitarity parameter $\eta_{\mu\tau}$ as a function of the
baseline length $L$. Solid curves denote the parent muon energy
$E_{\mu} =50~{\rm GeV}$ with CP phases being labeled in the figure,
while the dashed curve corresponds to $E_{\mu} =25~{\rm GeV}$ and
$\delta_{\mu\tau} =0$. The bounds on $|\eta_{\mu\tau}|$ in the MISS
for the NH and IH cases are also shown by red lines.} \vspace{-0.cm}
\end{center}
\end{figure}
In Fig.~\ref{fig:NF}, we display the sensitivity to $\eta_{\mu\tau}$
as a function of the baseline length $L$ for the near detector. One
can observe that such a setup provides indeed an excellent probe for
this type of non-unitarity effects. An interesting feature, which
appears if $\delta_{\mu\tau}$ is sizeable, is that the sensitivity
of the near detector will be improved around the baseline length
$L\sim 300 ~ {\rm km}$. This sensitivity enhancement could mainly be
regarded as a compromise between new physics effects and the
standard neutrino oscillation behavior. At a very short distance,
the transition probability $P_{\mu\tau}$ is determined by the last
term of Eq.~\eqref{eq:Pmutau}, whereas with increasing $L$, the
second term gradually dominates the flavor transitions. Thus, a
distance $L \lesssim 500~ {\rm km}$ (i.e., the CERN-Fr\'ejus
distance) would be favorable for the near detector.

\section{Summary}
\label{sec:summary}

We have proposed the MISS --- an economical low-scale seesaw
scenario with minimal particle content in the framework of the
inverse seesaw model. Compared to the generic inverse seesaw
mechanism, only two pairs of SM gauge singlets are introduced into
the MISS, which gives rise to strong correlations among the
non-unitarity parameters. Since one light neutrino has to be
massless in this scenario, we have discussed the experimental
constraints on these non-unitarity parameters in both NH and IH
cases. In view of our numerical and analytical results, the only
possibly sizable and phenomenologically interesting non-unitarity
parameter is $\eta_{\mu\tau}$, and the current upper bound on
$|\eta_{\mu\tau}|$ is improved from $1.1\times 10^{-3}$ to
$7.9\times 10^{-4}$ in the IH case. The Jarlskog invariants in the
presence of non-unitary neutrino mixing have been calculated. The
relative CP-violating phase of $\eta_{\mu\tau}$ is well constrained
by the structure of the MISS, and there are essentially no
observable CP-violating effects induced by $\delta_{\mu\tau}$ in the
IH case. We have also shown that the CP-violating effects emerging
in the MISS can be well tested at a future neutrino factory with an
OPERA-like near detector at a distance less than a few hundred
kilometers. The possible collider signatures at the LHC and the LFV
processes are also promising. However, a detailed analysis exceeds
the scope of the current work, and hence, it will be elaborated on
elsewhere.

Finally, we would like to stress that the MISS is motivated not only
by its simplicity and predictivity, but it can in particular be
viewed as a limit of the ``standard'' ISS setting when the heaviest
pseudo-Dirac neutrino essentially decouples. This is the kind of
behavior one would expect in grand unified models where a hierarchy
in the RH sector is often natural because of the link of the
RH-triplet Yukawa couplings to the other parts of the Yukawa
sector.\footnote{In this respect, it is worth noting that the
smallness of the RH neutrino mass scale does not in general obstacle
the grand-unified constructions, see for instance
Refs.~\cite{Malinsky:2005bi,Bertolini:2009qj} and references
therein.} Alternatively, it is often realized in certain classes of
flavor symmetric models in which the RH neutrinos are assigned to
some two-dimensional representations of the flavor group, i.e., the
smallest group containing one-, two-, and three-dimensional
representations of the symmetric permutation group $S_4$
\cite{Bazzocchi:2008ej}.
One can also adopt a variant of the several strategies proposed to
govern the flavor structure of the original inverse seesaw
framework, in particular, to accommodate the tri-bimaximal mixing
pattern, see e.g. Ref.~\cite{Hirsch:2009mx} and references therein.
However, a thorough implementation of a favor symmetry in the given
context is beyond the scope of the current work, and will not be
further discussed here.

\begin{acknowledgments}
\vspace{-2.5mm} We wish to thank Mattias Blennow and Enrique
Fern{\'a}ndez-Mart{\'i}nez for helpful discussions. We acknowledge
the hospitality and support from the NORDITA scientific program
``Astroparticle Physics --- A Pathfinder to New Physics'', March 30
- April 30, 2009 during which most of this study was performed. This
work was supported by the Royal Swedish Academy of Sciences (KVA)
[T.O.], the G{\"o}ran Gustafsson Foundation [H.Z.], the Royal
Institute of Technology (KTH), contract no.~SII-56510 [M.M.], the
Swedish Research Council (Vetenskapsr{\aa}det), contract
no.~621-2008-4210 [T.O.], and the National Natural Science
Foundation of China under grant nos. 10425522 and 10875131 [Z.Z.X.].
\end{acknowledgments}

\appendix

\section{Calculation of Jarlskog invariants}
\label{app}

The Jarlskog invariants for a non-unitary lepton flavor mixing
matrix $N$ are defined by
\begin{eqnarray}
\label{eq:A1} J_{\alpha\beta}^{ij}={\rm Im}(N_{\alpha i}N_{\beta
j}N^{*}_{\alpha j}N^{*}_{\beta i})\ ,
\end{eqnarray}
provided $N=(1-\eta)U$, where $U$ and $\eta$ are $3\times 3$ unitary
and Hermitian matrices, respectively. The stringent experimental
constraints on the deviation of $N$ from $U$ allow one to perform an
expansion of Eq.~\eqref{eq:A1} in powers of the small parameters
$\eta_{\alpha\beta}$ and $\theta_{13}$. Up to the second order in
$\eta_{\alpha\beta}$ and $s_{13}$, one has $ \label{eq:A2}
J_{\alpha\beta}^{ij} \simeq J +  \Delta J_{\alpha\beta}^{ij} \ , $
where $J=c_{12} c_{13}^2 c_{23} s_{12} s_{13} s_{23} \sin\delta$ and
\begin{eqnarray}
\label{eq:A3} \Delta J_{\alpha\beta}^{ij}  & = & - \sum_{\gamma}
{\rm Im}\left( \eta_{\alpha\gamma} U_{\gamma i} U_{\beta j}
U^*_{\alpha j} U^*_{\beta i} +  \eta_{\beta\gamma} U_{\alpha i}
U_{\gamma j}
U^*_{\alpha j} U^*_{\beta i}  \right. \nonumber \\
& + & \left. \eta^*_{\alpha\gamma} U_{\alpha i} U_{\beta j}
U^*_{\gamma j} U^*_{\beta i} + \eta^*_{\beta\gamma} U_{\alpha i}
U_{\beta j} U^*_{\alpha j} U^*_{\gamma i}  \right)\ .
\end{eqnarray}
In the parametrization \eqref{eq:para}, the nine relevant Jarlskog
invariants read:
\begin{eqnarray}\label{eq:A4}
\Delta J_{e\mu}^{12}  & = & - |\eta_{e\mu}| s_{12} c_{12} c_{23}
(1+c^2_{23}) \sin\delta_{e\mu} + |\eta_{e\tau}| s_{12} c_{12} s_{23}
c^2_{23} \sin\delta_{e\tau} \ , \\ \Delta J_{e\mu}^{23}  & = &
|\eta_{e\mu}| s_{12} c_{12} s^2_{23}c_{23}  \sin\delta_{e\mu} +
|\eta_{e\tau}| s_{12} c_{12}
s_{23}c^2_{23}  \sin\delta_{e\tau} \ , \\
\Delta J_{\mu\tau}^{23}  & = & -\Delta J_{e\mu}^{23} -
2|\eta_{\mu\tau}|  c^2_{12} s_{23}c_{23}  \sin\delta_{\mu\tau}\ , \\
\Delta J_{\mu\tau}^{31}  & = & -\Delta J_{e\mu}^{23} +
2|\eta_{\mu\tau}|  s^2_{12} s_{23}c_{23}  \sin\delta_{\mu\tau}\ , \\
\Delta J_{\tau e}^{12}  & = & |\eta_{e\mu}| s_{12} c_{12}
s^2_{23}c_{23}  \sin\delta_{e\mu} - |\eta_{e\tau}| s_{12} c_{12}
s_{23}(1+s^2_{23})  \sin\delta_{e\tau}\ , \\
\Delta J_{e\mu}^{23}  & = & \Delta J_{e\mu}^{31} = - \Delta
J_{\mu\tau}^{12} = \Delta J_{\tau e}^{23} = \Delta J_{\tau e}^{31} \
.
\end{eqnarray}


\end{document}